\documentclass[epsfig,12pt]{article}

\usepackage{epsfig}

\newcommand{\be}{\begin{equation}}
\newcommand{\ee}{\end{equation}}
\newcommand{\ba}{\begin{eqnarray}}
\newcommand{\ea}{\end{eqnarray}}
\newcommand{\bd}{\begin{displaymath}}
\newcommand{\ed}{\end{displaymath}}

\def\oneth{{\textstyle{\frac{1}{3}}}}

\title{
{\bf Critical Points in the Linear Sigma Model with Quarks}}
\author{{E. S. Bowman and J. I. Kapusta} \vspace*{0.1in}\\
{\it School of Physics and Astronomy, University of Minnesota}\\
 {\it Minneapolis, Minnesota 55455, USA}}

\date{September 30, 2008}

\begin{document}

\maketitle

\begin{abstract}
We employ a simple effective model to study the chiral dynamics of two flavors of quarks at finite temperature and density.  In particular, we determine the phase diagram in the plane of temperature and baryon chemical potential as a function of the pion mass.  An interesting phase structure occurs which results in zero, one or two critical points depending on the value of the vacuum pion mass.  
\end{abstract}

\vspace{0.5cm} {PACS numbers: 21.65.Qr, 25.75.Nq}

\newpage

\section{Introduction}

The physical pion mass is small but not zero.  In consequence, the conventional wisdom is that there is no true thermodynamic chiral phase transition at finite temperature $T$ and zero baryon chemical potential $\mu$, at least when the effects of the strange quark are neglected.  Instead, there is expected to be a curve of first order phase transition in the $\mu-T$ plane which terminates in a second order phase transition at some critical point $(\mu_c,T_c)$.  The location of the critical point obviously depends on the physical (vacuum) pion mass.  This topic has been under intense theoretical study using various effective field theory models, such as the Namu Jona-Lasinio model \cite{asakawa89,berges98,scavenius01}, a composite operator model \cite{barducci}, a random matrix model \cite{halasz98}, a linear sigma model \cite{scavenius01}, an effective potential model \cite{hatta02}, and a hadronic bootstrap model \cite{antoniou02}, as well as various implementations of lattice QCD \cite{fodor02,ejiri03,forcrand03,gavai05}.  Reviews of the subject were presented by Stephanov in the last few years \cite{stephanov}.  It is also of great interest because collisions between heavy nuclei at medium to high energy, such as at FAIR (Facility for Antiproton and Ion Research), which is under construction, may provide experimental information on the phase diagram in the vicinity of a critical point.  In contrast, current experiments at RHIC (Relativistic Heavy Ion Collider) and future experiments at LHC (Large Hadron Collider) may create too much entropy and therefore miss the high baryon densities needed to explore that region.  However, that depends on the location of the critical point.  In this paper we study the phase diagram of the linear sigma model coupled to two flavors of identical mass quarks.  Whereas the sigma model is an oft-used effective model that represents some of the essential features of the chiral dynamics of QCD, the reason to couple the fields to quarks is less obvious.  Why not couple to nucleons, delta resonances, etc. instead?  One argument is based on the existence of the critical point itself.  If a critical point exists, then one can go around it without crossing the curve of first order phase transition.  The effective degrees of freedom should not change too much in following such a path. Therefore, if quarks are considered to be reasonably useful degrees of freedom on the higher temperature side then they should be useful on the lower temperature side too.  One could make the same argument for using nucleons, delta resonances, and so on.  Presumably one could get the same answers.  However, it would entail including all the baryons in the Particle Data Tables and a multitude of coupling constants.  For this reason we use quarks to carry the baryon number but acknowledge the resulting uncertainty in the results.

The present work extends that of reference \cite{scavenius01} in two significant ways.  First, reference \cite{scavenius01} studies the linear sigma model with quarks but only in the mean field approximation.  The present work goes beyond that by including thermal fluctuations of the meson and fermion fields, which can be important at finite temperature when the magnitude of the fluctuations becomes comparable to or greater than the mean values. The technique we use was developed in several papers \cite{fraser,carter97,carter00}.  Second, reference \cite{scavenius01} fixed the pion mass at its vacuum physical value, whereas the present work does a scan of pion masses from zero to over 300 MeV, which may be particularly useful for comparison with lattice gauge theory with different quark masses.  The present work also extends that of reference \cite{Mocsy2004} to nonzero chemical potential, otherwise the techniques are the same.  Even when all other parameters of the model are fixed, we find an interesting phase diagram which may have zero, one or two critical points depending on the value of the vacuum pion mass.  

\section{The Linear Sigma Model with Quarks}

The Lagrangian is
\be
{\cal L} = \frac{1}{2}\left(\partial_{\mu}\mbox{\boldmath $\pi$}\right)^2
+ \frac{1}{2}\left(\partial_{\mu}\sigma\right)^2 - 
U(\sigma,\mbox{\boldmath $\pi$}) 
+ \bar{\psi} \left[ i \! \not\!\partial -
g \left( \sigma + i \gamma_5
\mbox{\boldmath $\tau$} \cdot \mbox{\boldmath $\pi$} \right) \right] \psi 
\ee
where
\be
U(\sigma,\mbox{\boldmath $\pi$}) = \frac{\lambda}{4}\left( \sigma^2+ \mbox{\boldmath $\pi$}^2 - f^2 \right)^2 - H \sigma 
\ee
in an obvious notation.  The $SU(2)_L\times SU(2)_R$ chiral symmetry is explicitly broken by the term $H \sigma$ which gives the pion a mass.  The scalar field has a nonvanishing vacuum expectation value $v$ determined at the classical level by the equation
\be
\lambda v \left( v^2 - f^2 \right) = H \, .
\ee
The scalar field is thus represented by a condensate plus a fluctuation
$\sigma = v + \Delta$.  The quarks have no intrinsic mass and only acquire one because of the condensate: $m_q = g v$.  The four parameters in the Lagrangian, $f$, $H$, $\lambda$ and $g$ are constrained by fixing the pion decay constant $f_{\pi} = 92.4$ MeV and the $\sigma$ mass $m_{\sigma} = 700$ MeV, while the quark mass is set to one-third of the nucleon mass $m_q = 313$ MeV.  The last required piece of information is the vacuum pion mass $m_{\pi}$, which is varied from 0 to $m_{\sigma}/2 = 350$ MeV.  At the classical level there are the simple relations
\ba
H &=& f_{\pi} m_{\pi}^2 \nonumber \\
\lambda &=& \frac{m_{\sigma}^2-m_{\pi}^2}{2f_{\pi}^2} \nonumber \\
f^2 &=& \frac{m_{\sigma}^2-3m_{\pi}^2}{m_{\sigma}^2-m_{\pi}^2}f_{\pi}^2 \nonumber \\
g &=& \frac{m_q}{f_{\pi}} \, .
\ea
All of this is standard practice, with the specific choices of $f_{\pi}$, $m_{\sigma}$ and $m_q$ corresponding to \cite{Mocsy2004}.  The value of the pion mass in the latter paper was fixed at its observed vacuum value of about 138 MeV.

\section{Self-Consistent Method for Determining the Equation of State}

The grand canonical partition function is
\be
{\cal Z}=\int[d\sigma][d\mbox{\boldmath $\pi$}]
[d\bar{\psi}][d\psi]\exp\left\{\int\limits_0^\beta d\tau\int\limits_V d^3x({\cal L}+ \oneth \mu\bar{\psi}\gamma^0 \psi)\right\}
\ee
where $\beta=1/T$ is the inverse temperature, $\mu/3$ is the quark chemical potential, $V$ is the volume of the system, and $\tau$ denotes imaginary time.
In order to create an effective mesonic model, we integrate out the quark degrees of freedom in the usual way \cite{kap}, such that
\be
\ln {\cal Z}_{\rm quark}=\ln \det {\it D}
\ee
where ${\it D}$ is the inverse quark propagator.  This, of course, still depends on the mesonic fields via an effective quark mass $m$, where 
\be
m^2=g^2(\sigma^2+\mbox{\boldmath $\pi$}^2)
\label{qmass}
\ee
which enters into the quasi-particle quark energy $E = \sqrt{p^2 +m^2}$.  This allows us to rewrite the grand canonical partition function solely in terms of mesonic degrees of freedom.  It amounts to the effective Lagrangian
\be
{\cal L} = \frac{1}{2}\left(\partial_{\mu}\mbox{\boldmath $\pi$}\right)^2
+ \frac{1}{2}\left(\partial_{\mu}\sigma\right)^2 -
U_{\rm eff}(\sigma,\mbox{\boldmath $\pi$})
\label{effL} 
\ee
where
\be
U_{\rm eff}(\sigma,\mbox{\boldmath $\pi$}) =
U(\sigma,\mbox{\boldmath $\pi$})
- \frac{T}{V} \, \ln {\cal Z}_{\rm quark}(\sigma,\mbox{\boldmath $\pi$})
\label{effU} 
\ee
is the effective potential.  The latter depends on both $T$ and $\mu$ as well as the mesonic fields.

The four equations of motion that follow from the effective mesonic Lagrangian eq. (\ref{effL}) are
\ba
\partial^\mu\partial_\mu\sigma + \frac{\partial U_{\rm eff}}{\partial\sigma}
&=& 0 \\
\partial^\mu\partial_\mu\pi_i+\frac{\partial U_{\rm eff}}{\partial\pi_i} &=& 0
\ea
with $i = 1, 2, 3$.  Now we decompose the mesonic fields into a condensate and fluctuations, as mentioned earlier.  We then write out the equations of motion in their formal series expansion in the fields and linearize them by making the replacement $\Delta^n \rightarrow n\langle\Delta^{n-1}\rangle\Delta$ where the angular brackets denote the ensemble average, see below.  We make use of the fact that, in this approximation, the sigma and pion field fluctuations are independent of each other, that is, $\langle\Delta^n\pi_i^r\rangle=\langle\Delta^n\rangle\langle\pi_i^r\rangle$.  The condition for the condensate is
\be
\left\langle\frac{\partial U_{\rm eff}}{\partial v}\right\rangle=0 \, .
\label{eomv}
\ee
The quasi-particle dispersion relations for the mesons, namely $E_\sigma^2=p^2+m_\sigma^2$ and $E_\pi^2=p^2+m_\pi^2$, involve effective masses given by
\ba
m_\sigma^2 = \left\langle
\frac{\partial^2 U_{\rm eff}}{\partial\Delta^2} \right\rangle \, ,
\label{esme}\\
m_\pi^2 = \left\langle
\frac{\partial^2 U_{\rm eff}}{\partial\pi_i^2}\right\rangle \, .
\label{epme}
\ea
The thermodynamic potential is then computed from
\be
\Omega=\langle U_{\rm eff}\rangle-\frac{1}{2}m_\sigma^2\langle\Delta^2\rangle
-\frac{1}{2}m_\pi^2\langle\mbox{\boldmath $\pi$}^2\rangle+\Omega_\sigma+\Omega_\pi
\ee
where $\Omega_\sigma$ and $\Omega_\pi$ are the independent particle contributions from the sigma and pion quasi-particles.  Explicitly
\be
\Omega_\sigma=\frac{T}{2\pi^2}\int\limits_0^\infty dp p^2\left[\frac{1}{2}\beta E_\sigma+\ln(1-e^{-\beta E_\sigma})\right]
\label{pot_sigma}\\
\ee
with a similar expression for the pions.

In this paper we will drop the shift in the zero point energy for simplicity.  Phenomenologically it may not make sense to keep it since it corresponds to high energy/short distance contributions which are probably not described well in this effective model.  The nontrivial issue of regularization and renormalization in self-consistent approximations has been discussed for the linear sigma model several times \cite{lenaghan00,verschelde01,vanhees02}.

Ensemble averaging is performed via the technique used in \cite{carter00,Mocsy2004}, which in turn is based on the analysis in \cite{carter97}; see also \cite{fraser}.  Consider some arbitrary functional $F(\sigma,\mbox{\boldmath $\pi$}^2)$, such as 
$\ln {\cal Z}_{\rm quark}(\sigma,\mbox{\boldmath $\pi$})$.  Expand this functional about the average values of the fields, namely $\sigma=v$ and $\mbox{\boldmath $\pi$}=\mbox{\boldmath $0$}$, and average term by term.
\begin{equation}
\langle F(\sigma,\mbox{\boldmath $\pi$})\rangle=\sum\limits_{k,n=0}^\infty F^{(k,n)}(v,\mbox{\boldmath $0$})\left\langle\frac{\Delta^k}{k!}\frac{\mbox{\boldmath $\pi$}^{2n}}{n!}\right\rangle
\label{expand}
\ee
Here 
\be
F^{(k,n)}(a,b)=\frac{\partial^{k+n}}{\partial a^k\partial b^n}F(a,b) \, .
\ee
We then need to relate $\langle\Delta^n\mbox{\boldmath $\pi$}^k\rangle$ to powers of $\langle\Delta^2\rangle$ and $\langle\mbox{\boldmath $\pi$}^2\rangle$.  Such relations were derived in \cite{carter97}.  We have $\langle\Delta^n\rangle=0$ for odd $n$ and $\langle\Delta^n\rangle=(n-1)!!\langle\Delta^2\rangle^{n/2}$ for even $n$.  For the pions, all the species are equivalent, so that $\langle\pi_1^2\rangle=\langle\pi_2^2\rangle=\langle\pi_3^2\rangle=\frac{1}{3}\langle\mbox{\boldmath $\pi$}^2\rangle$ and $\langle\mbox{\boldmath $\pi$}^{2k}\rangle=(2k+1)!!\langle\frac{1}{3}\mbox{\boldmath $\pi$}^2\rangle^k$.  After substituting back into (\ref{expand}), one notices that the averaging is equivalent to an integration over a Gaussian distribution, such that 
\be
\langle F(\sigma,\mbox{\boldmath $\pi$}^2)\rangle=\int\limits_{-\infty}^\infty dz P_\sigma(z)\int\limits_0^\infty dy y^2 P_\pi(y)F(v+z,y^2)
\label{av}
\ee
where 
\ba
P_\sigma(z)=\frac{1}{\sqrt{2\pi\langle\Delta^2\rangle}}\exp\left(-\frac{z^2}{2\langle\Delta^2\rangle}\right)\\
P_\pi(y)=\sqrt{\frac{2}{\pi}}\left(\frac{3}{\langle\mbox{\boldmath $\pi$}^2\rangle}\right)^{\frac{3}{2}}\exp\left(-\frac{3y^2}{2\langle\mbox{\boldmath $\pi$}^2\rangle}\right)
\ea
Note that in the limit of vanishing mean square fluctuations $\langle F(\sigma, \mbox{\boldmath $\pi$}^2)\rangle\rightarrow F(v,\mbox{\boldmath $0$})$, as one would expect.

It is important to know that the approximation employed here is thermodynamically self-consistent.  There are two consistency relations between the mesonic masses and fluctuations that follow directly, namely
\be
\langle\Delta^2\rangle=2\frac{\partial\Omega_\sigma}{\partial m_\sigma^2}
=\frac{1}{2\pi^2}\int\limits_0^\infty dp\frac{p^2}{E_\sigma}
\frac{1}{e^{\beta E_\sigma}-1}
\ee
and
\be
\langle\mbox{\boldmath $\pi$}^2\rangle=2\frac{\partial\Omega_\pi}{\partial m_\pi^2} = \frac{3}{2\pi^2}\int\limits_0^\infty dp\frac{p^2}{E_\pi}\frac{1}{e^{\beta E_\pi}-1} \, .
\ee
It is quite helpful to make use of a generalization of an expression in \cite{carter00} for the derivative of (\ref{av}) with respect to some parameter $\alpha$. After two integrations by parts, one obtains
\be
\frac{\partial}{\partial\alpha}\left\langle F(\sigma,\mbox{\boldmath $\pi$}^2)\right\rangle = \frac{\partial v}{\partial\alpha}\left\langle\frac{\partial F}{\partial v}\right\rangle+\frac{1}{2}\frac{\partial \langle\Delta^2\rangle}{\partial\alpha}\left\langle\frac{\partial^2 F}{\partial\Delta^2}\right\rangle
+\frac{1}{2} \sum_i \frac{\partial \langle \pi_i^2 \rangle}{\partial\alpha}\left\langle\frac{\partial^2 F}{\partial\pi_i^2}\right\rangle \, .
\ee
We can use this to calculate the derivative of the thermodynamic potential with respect to the mean field as well as the sigma and pion quasi-particle masses.  This gives
\bd
\frac{\partial\Omega}{\partial v} = \left\langle
\frac{\partial U_{\rm eff}}{\partial v}\right\rangle +\frac{1}{2}\frac{\partial\langle\Delta^2\rangle}
{\partial v} \left\{\left\langle\frac{\partial^2 U_{\rm eff}}
{\partial\Delta^2}\right\rangle-m_\sigma^2\right\} + 
\frac{1}{2}\frac{\partial\langle \pi_i^2\rangle}
{\partial v}\left\{\left\langle\frac{\partial^2 U_{\rm eff}}
{\partial\pi_i^2}\right\rangle-m_\pi^2\right\}=0
\ed
\bd
\frac{\partial\Omega}{\partial m_\sigma^2} = \frac{1}{2}\frac{\partial\langle\Delta^2\rangle}{\partial m_\sigma^2}\left\{\left\langle\frac{\partial^2 U_{\rm eff}}{\partial\Delta^2}\right\rangle-m_\sigma^2\right\}=0
\ed
\be
\frac{\partial\Omega}{\partial m_\pi^2} = \frac{1}{2} \sum_i
\frac{\partial\langle \pi_i^2\rangle}{\partial m_\pi^2}\left\{\left\langle\frac{\partial^2 
U_{\rm eff}}{\partial\pi_i^2}\right\rangle-m_\pi^2\right\}=0
\ee
These three derivatives vanish because of the averaged equation of motion and the two effective mass equations.  This gives us the thermodynamic consistency we require.

Due to thermodynamic consistency, we can calculate the energy density in the standard way, yielding
\ba
\epsilon &=& \langle U\rangle-\frac{1}{2}m_\sigma^2\langle\Delta^2\rangle-\frac{1}{2}m_\pi^2\langle\mbox{\boldmath $\pi$}^2\rangle+\frac{1}{2\pi^2}\int\limits_0^\infty dp\,p^2\left[ \frac{E_\sigma}{e^{\beta E_\sigma}-1}+3\frac{E_\pi}{e^{\beta E_\pi}-1}\right]\nonumber\\
&&+\frac{6}{\pi^2}\int\limits_0^\infty dp\,p^2 \left\langle \frac{E}
{e^{\beta(E-\mu)}+1}+\frac{E}{e^{\beta(E+\mu)}+1}\right\rangle \, .
\ea
The baryon number density can be calculated as
\be
n_B=\frac{2}{\pi^2}\int\limits_0^\infty dp\,p^2
\left\langle \frac{1}{e^{\beta(E-\mu)}+1}-
\frac{1}{e^{\beta(E+\mu)}+1}\right\rangle \, .
\ee
Note that in the quark terms there are pairs of angular brackets indicating the averaging procedure of eq. (\ref{expand}).  This means that the momentum integrals are evaluated for a particular configuration of the mesonic fields with quark mass given by eq. (\ref{qmass}), and then the mesonic fields are averaged over.  One may choose to define a thermally averaged effective quark mass by the ensemble average of eq. (\ref{qmass}), to wit
\be
m^2 \rightarrow m^2_q = g^2 \left( \langle \sigma^2 \rangle + 
\langle \mbox{\boldmath $\pi$}^2 \rangle \right) =
g^2 \left( v^2 + \langle \Delta^2 \rangle + 
\langle \mbox{\boldmath $\pi$}^2 \rangle \right) \, .
\label{effqmass}
\ee
However, the quark integrals are not evaluated with this single mass.

\section{Numerical Results}

The equation of motion and the effective meson mass equations must be solved self-consistently.  The integrals in the thermal averages are calculated using an adaptive $n$-point rule algorithm, and the thermodynamic integrals are calculated using an adaptive Simpson's rule algorithm.  We have verified numerically in a representative sample of combinations of $T$ and $\mu$ that the thermodynamic identities
\be
\epsilon(\mu,T) = -P(\mu,T) + T \frac{\partial P(\mu,T)}{\partial T} 
+ \mu n_B(\mu,T)
\ee
and
\be
n_B(\mu,T) = \frac{\partial P(\mu,T)}{\partial \mu}
\ee
are satisfied.  For more details on the mathematical derivations and numerical procedures see \cite{BowmanPhD}.

To determine the location of a phase transition, one computes the pressure as a function of temperature and chemical potential.  An example is shown in Fig. 1, where $P$ is plotted as a function of $\mu$ for two different temperatures.  In this example the vacuum pion mass is fixed at its physically observed value.  For the higher temperature of 80 MeV there is only one self-consistent solution to the equations described above.  For the lower temperature of 50 MeV, there is a unique solution at large $\mu$ and another unique solution at small $\mu$.  For a range of $\mu$ centered about 900 MeV there are three solutions: one is associated with a continuation of the low density phase, one is associated with a continuation with the high density phase, while the third solution (not shown in the figure) is an unstable phase.  The point where the two curves cross is the location of the phase transition.  In this example it is first order since there is a discontinuity in the slope $\partial P(\mu,T)/\partial \mu \equiv n_B$.  Where each curve terminates is the limit of metastability for that phase.  The thermodynamically favored phase is the one with the largest pressure.  At some temperature between 50 and 80 MeV the slopes are equal at the crossing point, there are no metastable phases, and the second derivative is discontinuous.  This is indicative of a second order phase transtion.  The location of this point in the $\mu-T$ plane is the critical point.  The critical point has coordinates designated $(\mu_c,T_c)$.

In Fig. 2 is plotted the energy density divided by $T^4$ versus $T$, a commonly used plot in studies of the equation of state at zero baryon chemical potential, for vacuum pion masses of 0 and 200 MeV.  For zero pion mass the phase transition is first order, whereas for a vacuum pion mass of 200 MeV there is no thermodynamic phase transition, only a crossover from one phase to the other.  The contribution to the energy density for a gas of massless bosons is $(\pi^2/30)T^4$ per degree of freedom, whereas massless fermions contribute $7/8$ of that amount due to the difference between Bose and Fermi statistics.  If all the degrees of freedom in this model were massless at high temperature the ratio $\epsilon/T^4$ would be about 8.22.  The results shown indicate a value of about 7.5 around a temperature of 200 MeV and not changing very quickly with increasing temperature.  Obviously this is due to interactions, in particular the quasi-particles acquiring effective masses at high temperature due to thermal fluctuations.  It should be remarked that if one considered a gas of massless quarks with 24 fermionic degrees of freedom (same as in this model) plus massless gluons with 16 bosonic degrees of freedom (versus 4 in this model) one would have $\epsilon/T^4$ approximately 12.17.  Calculations with lattice QCD just above $T = 200$ MeV yield a value about 3/4 as large \cite{latticeDOF}.  This indicates that the number of effective massless degrees of freedom is less, due to interactions, which is one argument for referring to the matter in this temperature range as strongly-coupled quark-gluon plasma.  The sigma model with quarks exhibits a much smaller decrease in the effective number of degrees of freedom, suggesting that the fewer degrees of freedom incorporated in the sigma and pion fields versus the gluon fields may be qualitatively reasonable.

We should also remark that when the vacuum pion mass is set to 138 MeV (curve not shown) we reproduce the numerical results of reference \cite{Mocsy2004}, a check on our (and their) numerical work.

The phase diagram in the $\mu-T$ plane for a sampling of vacuum pion masses is shown in Fig. 3.  For the sake of discussion, pick one, for example $m_{\pi,{\rm vac}} = 138$ MeV.  There is a curve of first order phase transition starting on the $\mu$ axis and arching to the left.  This curve terminates at a critical point of approximately $T_c = 75$ MeV, $\mu_c = 850$ MeV.  Representative curves for $m_{\pi,{\rm vac}}$ = 50 and 200 MeV are also shown.  For $m_{\pi,{\rm vac}}$ = 0 there is no critical point; the curve smoothly extends from the $\mu$ to the $T$ axis. For $m_{\pi,{\rm vac}} = 321$ MeV the critical point sits on the $T = 0$ axis, and for $m_{\pi,{\rm vac}} > 321$ MeV there is no phase transition at all.  Of course the precise numbers depend on the constants in this model, such as the vacuum sigma mass and what value one assigns to the coupling of the quark field to the sigma field, but the results are in line with expectations \cite{stephanov}.  In particular, reference \cite{scavenius01} found $T_c = 99$ MeV and $\mu_c = 621$ MeV using the physically observed pion mass.  Reference \cite{scavenius01} used the mean field approximation whereas the present work also includes thermal fluctuations in the meson and quark fields, and in addition that paper fixed the vacuum sigma mass at 600 MeV compared to the present value of 700 MeV.  The increase in $T_c$ and decrease in $\mu_c$ compared to our results is consistent with a sort of universal curve, as exemplified by Fig. 3.

Something very interesting happens as the vacuum pion mass is decreased from 50 MeV to 0.  Figure 4 shows the phase diagram for a vacuum pion mass of 35 MeV.  There are now two critical points!  There is a line of first order phase transition beginning on the $\mu$-axis and arching to the left to end at a critical point of $\mu_{c1} \approx 725$ Mev and $T_{c1} \approx 92$ MeV, and another line of first order phase transition beginning on the $T$-axis and arching to the right to end at a critical point of $\mu_{c2} \approx 240$ Mev and $T_{c2} \approx 137$ MeV.  In the vicinity of $\mu = 575$ MeV and $T = 110$ MeV the latent heat gets pinched to zero for some critical value of the vacuum pion mass between 0 and 35 MeV.  We have not attempted to pinpoint the exact value of the vacuum pion mass since it is quite numerically intensive and depends on all the other parameters in the model, as well as the structure of the model itself.

The latent heat along the critical curve is plotted against $T$ in Fig. 5 for vacuum pion masses of 0, 35, and 138 MeV.  At $T=0$, corresponding to large chemical potential, the latent heats are on the order of 270 MeV/fm$^3$.  Where they go to zero is a critical point.  Note the nonzero latent heat in the bottom right corner of the figure; it corresponds to the region in the upper left corner of Fig. 4.  In that region for that value of vacuum pion mass the latent heat is very small, less than about 10 MeV/fm$^3$.  For not much larger values of the vacuum pion mass this latent heat shrinks to zero. 

The results shown in Fig. 3 illustrate the conventional view on the nature and location of the critical point.  The result shown in Fig. 4 is unconventional or exotic.  To the best of our knowledge, this is the first time a model calculation has resulted in such a phase diagram.  It is well known that the nature of the phase transition or crossover for two flavors of quarks is sensitive to such details as the strength of the axial U(1) anomaly and the value of the vacuum $\sigma$ mass.  For example, see \cite{Pisarski,Lenaghan,Chandra}.  It cannot be expected that a model as simple as the one studied here can make definitive predictions for what actually happens in QCD with the observed values of the quark masses.  However, it can serve to illustrate the possibilities.

\section{Conclusion}

We have utilized a relatively simple yet self-consistent model to study the curve of chiral phase transitions in the plane of temperature and baryon chemical potential as a function of the vacuum pion mass.  For a range of vacuum pion mass from zero to some critical value, which for the parameters we have chosen in this paper is less than 35 MeV, there is a continuous curve of first order phase transition from the $T$ axis to the $\mu$ axis.  As the vacuum pion mass increases, two critical points emerge; this is an unconventional, or exotic picture.  Further increase in the pion mass causes one critical point to hit the $T$ axis and disappear while the other critical point moves to smaller $T$ and larger $\mu$.  For a pion mass of 321 MeV this remaining critical point hits the $\mu$ axis and also disappears.  Changing the other parameters in the model, such as the vacuum $\sigma$ mass or the strength of the coupling of the quarks to the chiral fields, will obviously change the particular values of the pion masses for which two, one or no critical points exist.  The existence and behavior of the critical point at large $T$ and small $\mu$ is consistent with the lattice calculations of de Forcrand and Philipsen \cite{forcrand03}.  (A recent preprint by Fukushima \cite{Kenji} shows that the Nambu Jona-Lasinio model with sufficiently strong vector coupling also supports a critical point at large $T$ and small $\mu$.)  The existence and behavior of the critical point at small $T$ and large $\mu$ is the conventional picture.  The model presented here shows that both pictures can coexist for a range of pion masses.  More details on the numerical solutions as well as results when the aforementioned parameters are varied will be presented elsewhere \cite{BowmanPhD}.

Further work with this and related models are easily identified.  It would be interesting to extend the sigma model to include strangeness, and to vary both the vacuum pion and kaon mass, using the self-consistent techniques of the present paper.  It should be straightforward to calculate shear and bulk viscosities and thermal conductivity along the lines of \cite{Paech} and \cite{Sasaki}.  These transport coefficients could be used in fluid dynamic modeling of heavy ion collisions at FAIR and for low energy runs at RHIC.  These projects are under current investigation.

\section*{Acknowledgements}

We thank E. E. Kolomeitsev for advice throughout the course of this work.  We also gratefully acknowledge Ph. de Forcrand for discussions and suggestions.  This work was supported by the US Department of Energy (DOE) under grant 
DE-FG02-87ER40328.

\newpage

\section*{Appendix}

To solve the equations of motion one needs the derivative of 
\be
U_{\rm quark} = -(T\ln {\cal Z}_{\rm quark})/V
\ee
with respect to $\Delta$
\be
\left\langle\frac{\partial U_{\rm quark}} {\partial\Delta}\right\rangle=g^2\langle\sigma A(m)\rangle
\ee
where the function $A$ is the fermionic analogue of the mesonic mean square fluctuations
\be
A(m) = 2\frac{\partial U_{\rm quark}}{\partial m^2} =\frac{6}{\pi^2}\int\limits_0^\infty dp\frac{p^2}{E}
\left[\frac{1}{e^{\beta(E-\mu)}+1}\nonumber\\
+ \frac{1}{e^{\beta(E+\mu)}+1}\right] \, .
\ee
Within the current approach the quark condensate can be expressed as
\be
\langle\bar{\psi}\psi\rangle=\frac{1}{g}
\left\langle\frac{\partial U_{\rm quark}}{\partial\sigma}\right\rangle
=g\langle\sigma A(m)\rangle \, .
\ee
One also needs the second derivatives
\bd
\left\langle\frac{\partial^2 U_{\rm quark}}{\partial\Delta^2}\right\rangle = g^2\left\langle A(m)+2g^2\sigma^2\frac{\partial A}{\partial m^2}\right\rangle
=\frac{g^2}{\langle\Delta^2\rangle}\langle\Delta(v+\Delta)A(m)\rangle
\ed
\be
\left\langle\frac{\partial^2 U_{\rm quark}}{\partial\pi_i^2}\right\rangle = g^2\left\langle A(m)+2g^2\pi_i^2\frac{\partial A}{\partial m^2} \right\rangle\nonumber\\
=\frac{g^2}{\langle\pi_i^2\rangle}\langle\pi_i^2A(m)\rangle
\ee
where the right hand sides are derived using the Hartree approximation in the series representation of the thermal averaging process.  The required derivative is
\be
\frac{\partial A}{\partial m^2} = -\frac{3}{\pi^2}\int\limits_0^\infty dp\frac{1}{E}\left[\frac{1}{e^{\beta(E-\mu)}+1}
+ \frac{1}{e^{\beta(E+\mu)}+1}\right] \, .
\ee

One also requires the averages of the mesonic potential and its derivatives.  For the pressure and energy density one needs the averaged potential itself.
\be
\langle U\rangle = \left\langle\frac{\lambda}{4}(\sigma+\mbox{\boldmath $\pi$}^2-f^2)^2-H\sigma\right\rangle
\ee
For the equation of motion one needs the first derivative of the bare potential
\be
\left\langle\frac{\partial U}{\partial\Delta}\right\rangle=\lambda v\left(v^2+3\langle\Delta^2\rangle+\langle\mbox{\boldmath $\pi$}^2\rangle-
f^2 \right)-H \, ,
\ee
and for the mesonic masses one needs the second derivatives
\ba
\left\langle\frac{\partial^2 U}{\partial\Delta^2}\right\rangle&=&\lambda \left(3v^2+3\langle\Delta^2\rangle+\langle\mbox{\boldmath $\pi$}^2\rangle
-f^2\right)\\
\left\langle\frac{\partial^2 U}{\partial\pi_i^2}\right\rangle&=&\lambda \left(v^2+\langle\Delta^2\rangle+\frac{5}{3}\langle\mbox{\boldmath $\pi$}^2\rangle-f^2\right) \, .
\ea

\begin{figure}
\begin{center}\includegraphics[width=5.0in,angle=0]{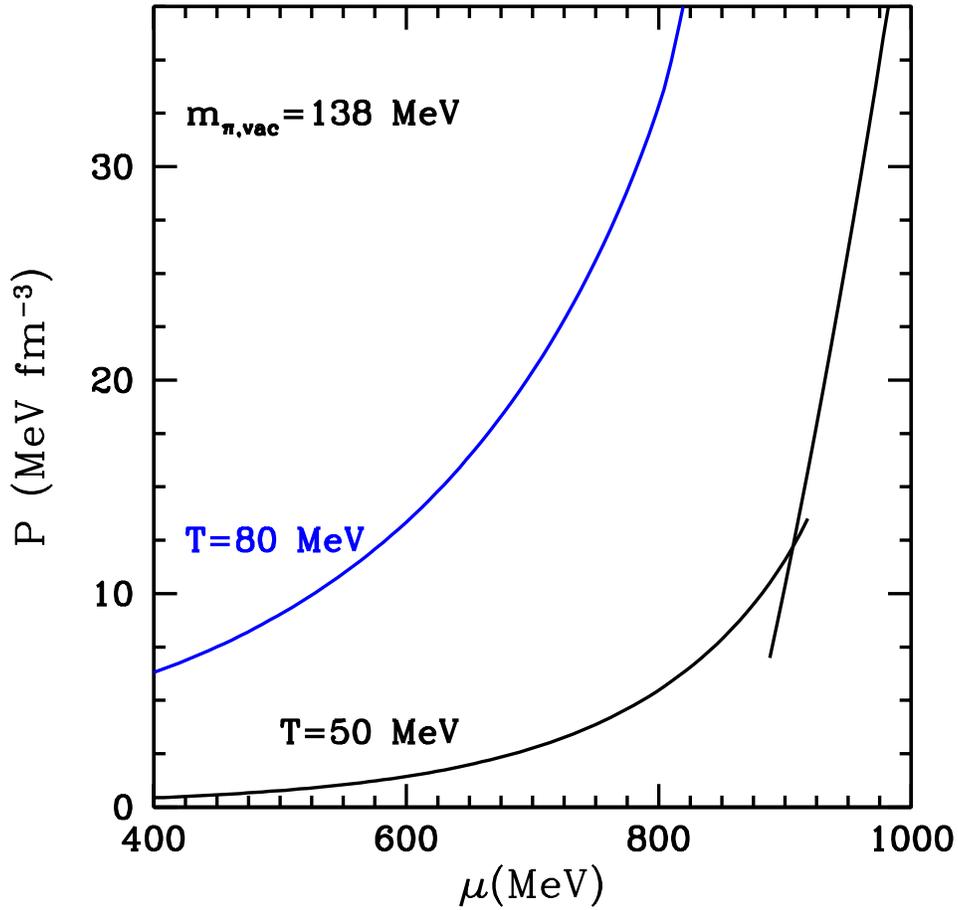}
\caption{Pressure versus baryon chemical potential for two different temperatures with the physical value of the pion mass in vacuum.  At the lower temperature there is a discontinuity in the slope at the crossing point, indicating a first order phase transition.  The curves are extended to the limits of metastability.  At the higher temperature the pressure is smooth and there is no phase transition.}
\end{center}
\label{pressures}
\end{figure}

\begin{figure}
\begin{center}\includegraphics[width=5.0in,angle=0]{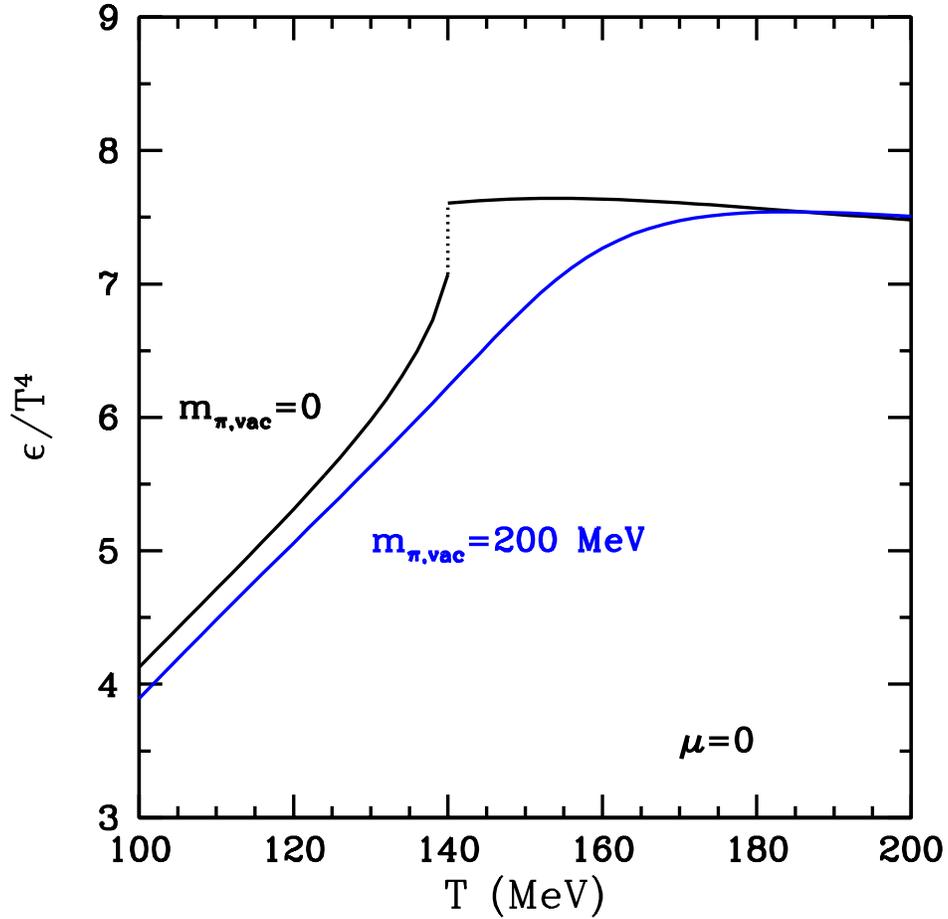}
\caption{The energy density, normalized to $T^4$, versus $T$ at zero chemical potential.  For zero vacuum pion mass there is a first order transition, whereas for a vacuum pion mass of 200 MeV there is only a rapid crossover.}
\end{center}
\label{energies}
\end{figure}

\begin{figure}
\begin{center}\includegraphics[width=5.0in,angle=0]{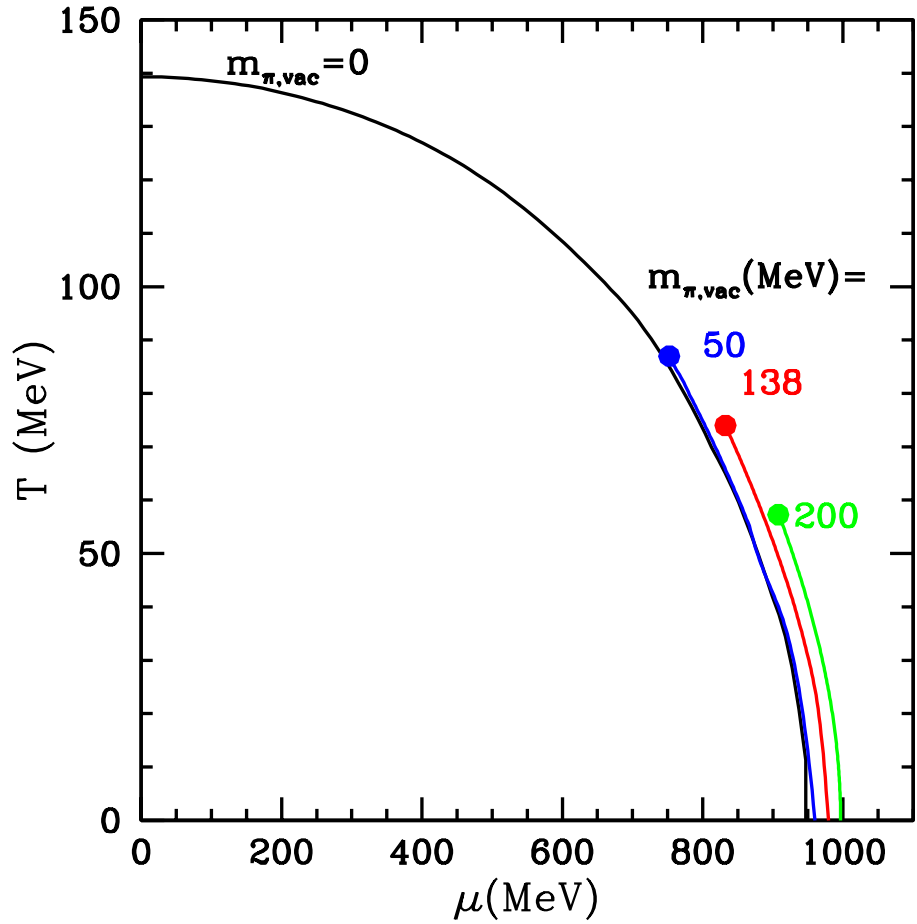}
\caption{Phase diagram of this model for illustrative values of the vacuum pion mass.  The solid curves represent a first order phase transition, terminating at a critical point where the transition is second order.  For zero vacuum pion mass the curve is first order all the way from one axis to the other.}
\end{center}
\label{transitionsall}
\end{figure}

\begin{figure}
\begin{center}\includegraphics[width=5.0in,angle=0]{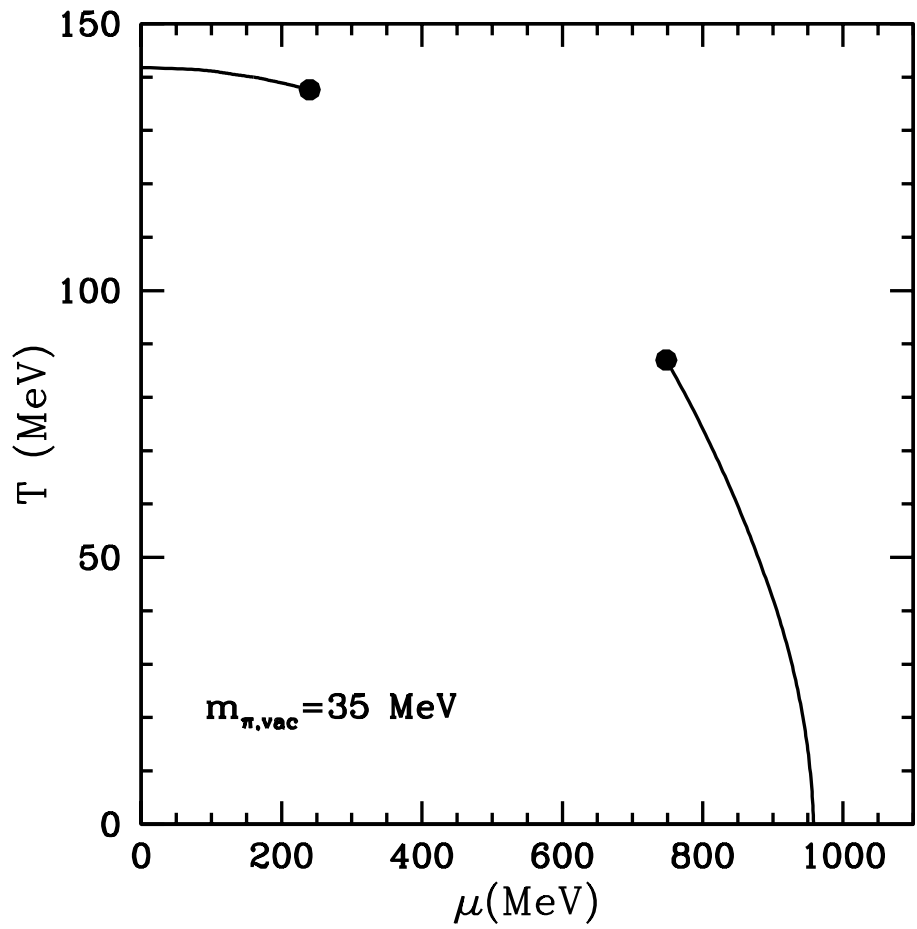}
\caption{Phase diagram of this model for a vacuum pion mass of 35 MeV.  The solid curves represent a first order phase transition, terminating at a critical point where the transition is second order.  For this value of the vacuum pion mass there are two critical points.}
\end{center}
\label{transition35}
\end{figure}

\begin{figure}
\begin{center}\includegraphics[width=5.0in,angle=0]{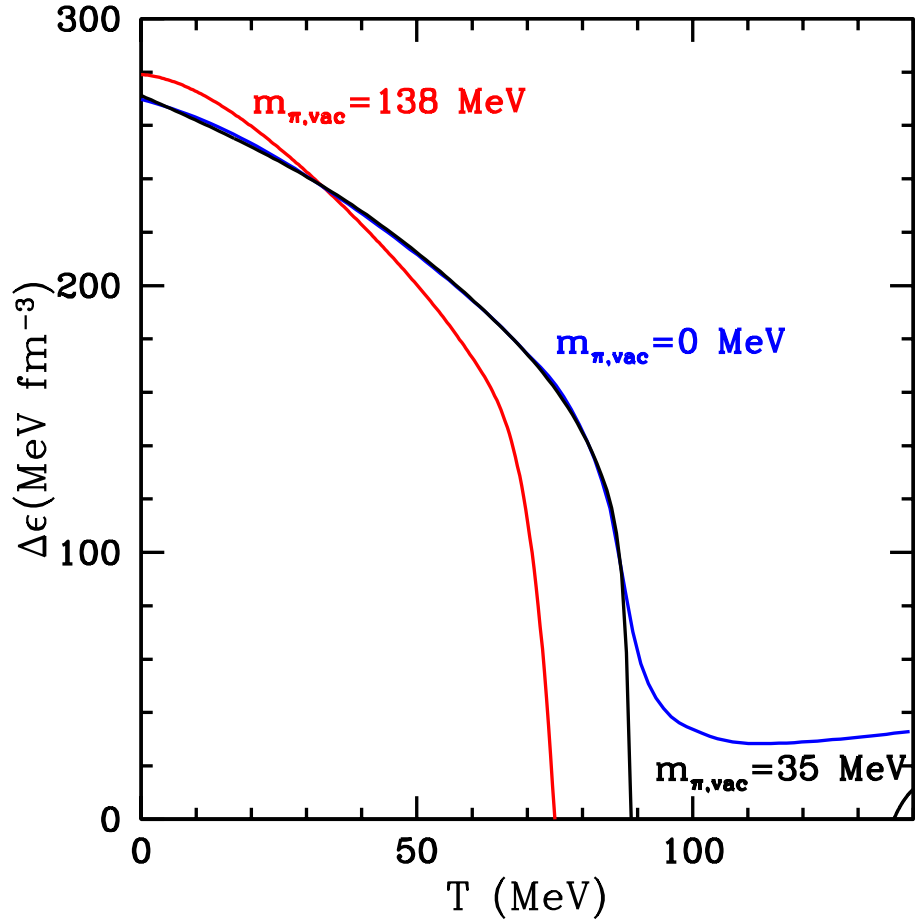}
\caption{Latent heat along the critical curve for various vacuum pion masses.  The critical point corresponds to the vanishing of the latent heat.  Note that for a vacuum pion mass of 35 MeV there is a second region in the bottom right corner.}
\label{latentheatall}
\end{center}
\end{figure}

\end{document}